\documentclass[sigconf]{acmart}

\usepackage{amsmath}
\usepackage{array}
\usepackage{amssymb}
\usepackage{enumitem}
\usepackage{textcomp}
\usepackage{gensymb}
\usepackage{siunitx}
\usepackage{rotating}
\usepackage{multirow}
\usepackage{xspace}
\usepackage{xcolor,color}
\usepackage{blkarray}
\usepackage{multirow}
\usepackage{graphicx}
\usepackage{epstopdf}
\usepackage{wrapfig}
\usepackage{xspace}
\usepackage[utf8]{inputenc}
\usepackage[font=small,skip=5pt]{caption}
\usepackage{subcaption}
\usepackage{mathtools}
\usepackage{mathptmx}
\usepackage{lmodern}
\usepackage{supertabular}
\usepackage{adjustbox}

\makeatletter
\def\@copyrightspace{\relax}
\makeatother

\begin{document}

\title{On the Relation Between Mobile Encounters and Web Traffic Patterns: A Data-driven Study}

\acmYear{2018}
\setcopyright{none}
\acmConference[Technical Report for MSWIM '18 conference paper]{21st ACM International Conference on Modelling, Analysis and Simulation of Wireless and Mobile Systems}{October 28-November 2, 2018}{Montreal, QC, Canada.}
\acmBooktitle{21st ACM International Conference on Modelling, Analysis and Simulation of Wireless and Mobile Systems (MSWIM '18), October 28-November 2, 2018, Montreal, QC, Canada}
\acmPrice{15.00}
\acmDOI{10.1145/3242102.3242137}
\acmISBN{978-1-4503-5960-3/18/10}
\settopmatter{printacmref=false}

\author{~Babak Alipour, Mimonah Al Qathrady, Ahmed Helmy}
\affiliation{%
  \institution{
  Department of Computer and Information Science and Engineering\\
University of Florida}
  \city{Gainesville}
  \state{FL, USA}
}
\email{{babak.ap , mimonah , helmy}@ufl.edu}

\thispagestyle{plain}
\pagestyle{plain}

\begin{abstract}

Mobility and network traffic have been traditionally studied separately. 
Their interaction is vital for generations of future mobile services and effective caching, but has not been studied in depth with real-world big data.
In this paper, we characterize mobility encounters and 
study the correlation between encounters and web traffic profiles using large-scale datasets (30TB in size) of WiFi and NetFlow traces. 
The analysis quantifies these correlations for the first time, across spatio-temporal dimensions, for device types grouped into on-the-go \textit{Flutes} and sit-to-use \textit{Cellos}. 
The results consistently show a clear relation between mobility encounters and traffic across different buildings over multiple days,
with encountered pairs showing higher traffic similarity than non-encountered pairs, and long encounters being associated with the highest similarity.
We also investigate the feasibility of learning encounters through web traffic profiles, with implications for dissemination protocols, and contact tracing.
This provides a compelling case to integrate both mobility and web traffic dimensions in future models,
not only at an individual level, but also at \textbf{\textit{pairwise}} and collective levels.
We have released samples of code and data used in this study on GitHub, to support reproducibility and encourage further research (\url{https://github.com/BabakAp/encounter-traffic}).

\end{abstract}

\maketitle

\section{Introduction \& Related work}

The effect of mobility and network traffic on wireless networks has been clearly established in the literature (e.g. \cite{Important}). Several efforts studied models of mobility and network traffic, albeit mostly separately and in isolation. 
There is a vast body of research focused on mobility or traffic independently, which we cannot possibly exhaustively cover.
We refer the reader to \cite{treuniet:csur14, hess:csur16} for surveys of mobility modeling and analysis.
Some of the most advanced studies on mobility \cite{Cobra} have identified individual \cite{TVC}, pairwise (encounter), and collective (group) dimensions for mobility modeling. That study, however, did not consider traffic.
We hope to bridge that gap by analyzing the interplay of mobility and traffic at the pairwise level.

We argue that the relation between mobility and traffic needs further in-depth analysis, as it will likely be the center of many future mobile services.
In this paper, we focus on the \textbf{\textit{pairwise} (encounter)} dimension of mobility and study its interplay with the traffic patterns of mobile users. 
Aside from this study's importance to realistic modeling, simulation and performance evaluation of next-generation networks, it is quite relevant to encounter-based services, e.g., content sharing, opportunistic networking, mobile social networks, and encounter-based trust and homophily (\cite{manweiler2009smile, dong2014inferring}), to name a few.

Encounters between mobile nodes have been studied in previous research (e.g., in \cite{enNodal, enMetric}) to characterize opportunities of inter-user encounters. Others (e.g., \cite{EncCollection1, EncCollection2}), mainly collect encounter traces using mobile devices to analyze, model and understand communication opportunities in different settings.
None of these studies, however, analyze traffic nor the correlation between encounters and real-world traffic patterns. 
In this study, we focus on the \textbf{interplay between traffic and encounters}, while considering the score (e.g., duration, frequency) of the encounter events. 
We use extensive data-driven analyses to quantify the correlations between network traffic and encounter scores.

Several studies analyzed wireless traffic flows \cite{FlowStudy}, and mobile web-visitation patterns \cite{Saeed}. These studies, however, did not investigate the relation with mobility and node encounters.

In addition, many research studies on mobility encounters or traffic patterns did not consider \emph{device type}.
Devices' form factor affects mode of usage, leading to varied traffic profiles 
(\cite{afanasyev2008analysis, maier2010first, Falaki2010, chen2012network, Kumar2013, gember2011comparative, papapanagiotou2012smartphones}).
But these studies do not study the interplay of traffic with mobility and encounters.
These devices are also used during different modes of transportation.
Smartphones and e-readers, which we refer to as \textit{\textbf{Flutes}}, are devices used \textbf{'on-the-go'}. On the other hand, laptops are \textbf{'sit-to-use'} devices and are referred to as \textit{\textbf{Cellos}} in this study.  
In our earlier work \cite{Alipour2018}, we contrast various \textit{mobility} and \textit{traffic} features of \textit{Flutes} and \textit{Cellos}, including radius of gyration, location visitation preferences,
and flow-level statistics.
But that study only investigated mobility and traffic features across the \textbf{\textit{individual}} dimension.
While we investigate the \textbf{\textit{pairwise}} dimension here, focusing on encounter-traffic interplay while considering the device types. 

We use extensive traces from our collected datasets (with $76$B records, and $\approx$30TB in size) covering over 78K devices, in 140 buildings on a university campus. 
The data includes information about WiFi associations, as well as DHCP and NetFlow traces, covering the dimensions of mobility and network traffic. 
The data is sanitized and categorized based on buildings, days, device types (\textit{Flutes}, \textit{Cellos}), and encounter duration, then the analysis is done across all these dimensions.

The main question addressed in this study is ‘\emph{How do device encounters affect network traffic patterns, across time, space, device type and encounter duration?}’ 
For that purpose, we: i- analyze mobility encounters patterns, ii- define web traffic profiles for users, and iii- look at their interplay.
Although this question has not been directly studied in-depth before, our findings are quite surprising, showing that for the majority of buildings a consistent correlation exists between traffic profiles and encountered (vs. non-encountered) pairs of users. 
We also found the correlation to be the strongest for \textit{Cello-Cello} encounters on weekends.
Further, we find that such relation strengthens for long encounters, while short encounters are not significantly different from non-encountered pairs.
Finally, we utilized a deep learning model to learn encounters of user pairs in a day and building based on their traffic profiles alone. The model achieved a high accuracy (90\%+) in many settings, with major implications for encounter-based services, rumor anatomy analysis, and infection tracing.

These results can potentially impact a variety of applications, 
including those utilizing prediction of traffic load/demand using encounters, and vice versa.
In addition, mobility modeling and protocol evaluation could benefit from deep integration of (and the interplay between) encounters and traffic.
We hope for this paper to be the first in a series of studies on mobility and traffic, and their interplay, across individual, \textit{pairwise} and collective (group) dimensions, towards fully \emph{integrated realistic traffic-mobility models}.

The rest of the document is organized as follows. 
Section 2 describes the datasets used in detail and their processing. 
Section 3 defines and analyzes mobility encounters. 
Section 4 introduces the web traffic profiles. 
Then, section \ref{sec:pairwise} presents the \textit{pairwise} encounter-traffic relationship. 
Next, section \ref{sec:dnn} introduces our encounter learning methods and summary of results.
Finally, Section \ref{sec:conclusion} discusses the findings, future work and concludes the paper.\\
\section{Datasets}

This study utilizes multi-sourced large-scale datasets we have collected including WLAN, DHCP, NetFlow, and other external sources (e.g., maps, rDNS).

\subsection{Wireless LAN (WLAN) \& Encounters}
The WLAN event logs were collected on a university campus during April 2012.
Each log entry provides a timestamp, an IP address at a corresponding
access point (AP) and MAC address of the associated user device.
There are 1,700 APs and $\approx78k$ devices in this dataset.
In this study, we analyze the device behavior, as identified by its MAC address\footnote{MAC address randomization was introduced on popular platforms after our traces were collected, and does not affect our association trace.}. 

Pairwise user mobility behavioral patterns are represented through the patterns of encounters between two mobile nodes.
An encounter is defined as when two user devices are associated with the same AP at an overlapping time interval. 
The \textit{\textbf{Encounter}} traces are generated based on WLAN logs.
An example of a pairwise encounter record, constructed from WLAN traces, is shown in Table \ref{table:encounterRecord}.

\subsection{Location Information}
To analyze traces in different places, location information of APs is required.
Since exact locations of the APs were not available, the APs are assigned 
approximate locations based on the building where they are installed, 
i.e. building latitude/longitude from Google Maps API.
The crowd-sourced service wigle.net was used to validate this positioning.
From 130 matched APs (7.6\% of total), in 58 buildings, 
all were within 200m or less from their mapped location. 
This error (1.5\% of campus area) is reasonable considering the maximum AP coverage range, 
inaccurate coarse-grained localization services and that we use
coordinates of the center of each building whereas users
may see an AP on the edge of a building.

\subsection{NetFlow}
The Netflow traces are collected from the same network, during April 2012\footnote{The last five days of NetFlow cover exam dates and are omitted, since those dates do not represent a typical campus environment.}.
A flow is a unidirectional sequence of packets transferred between two IP hosts, with each flow record
retaining over 30 fields,
including flow start/finish times, source/destination IP addresses and ports, transport protocol and traffic volume (packets/bytes).
In raw format, dataset size is $\approx30TB$, providing a vast, high granularity data source for $\approx78k$ devices. 
Due to quadratic asymptotic growth of \textit{pairwise} traffic analysis, for this study, we focus on the 10,000 most active users in terms of traffic consumption, to keep computations manageable.
Table \ref{table:netflow} provides an example flow with a subset of important features.

\subsection{Device Type Classification}
\label{datasets:devTypeClassification}

To classify devices into \textbf{'on-the-go' \textit{Flutes}} and \textbf{'sit-to-use' \textit{Cellos}}, 
we build upon the same observations and heuristics of our previous work \cite{Alipour2018}.
First, the device manufacturer can
be identified based on the first 3 octets of the MAC
address (Organizationally Unique Identifier).  
Most manufacturers produce one type of device (either
laptop or phone), but some produce both (e.g., Apple).  
In the latter case, OUI used for one device type is not used for another.  
To validate, a survey was carried out and 30 MAC prefixes were accurately classified.
OUI and survey information helped identify and classify 46\% of all devices.
Then, from the \textit{NetFlow} logs of these labeled devices, we observe over 3k devices (92\% of which are
\textit{flutes}) contacting \textit{admob.com}; an ad platform serving mainly
smartphones and tablets (i.e. flutes). 
This enables further classification of the remaining MAC addresses, with reasonable accuracy, using the following heuristic: 
(1) obtain all OUIs (MAC prefix) that contacted \textit{admob.com}; 
(2) if it is unlabeled, mark it as a flute.  
Overall, over 97\% of devices in NetFlow traces were labeled ($\approx50$K flutes and $\approx27$K cellos).

This enables classification of pairwise encounters, based on the encounter pair device types:
1) \textbf{\emph{Flute-Flute} (\emph{FF})}: encounter event between two flutes.
2) \textbf{\emph{Cello-Cello} (\emph{CC})}: the pair are cello devices.
3) \textbf{\emph{Flute-Cello} (\emph{FC})}: encounter event between a flute and cello.

\begin{table*}[h!]
\centering
\caption{Encounter record example}
\begin{adjustbox}{max width=\textwidth}
  \begin{tabular}{*{9}{|c}|}
\hline
User1 Mac& User2 Mac& User1 Asso. Start & User1 Assoc. End & Access Point Mac & User2 Asso. Start& User2 Association End & Encounter Start & Encounter End   \\

\hline
7c:61:93:9d:30:2e & cc:08:e0:34:a7:3e & 1334503199  & 11334503337 & 1334501153 & 1334506764 & bcftr0gb-win-lap1142-1 & 1334503199 & 1334503337 \\
\hline
\end{tabular}
\end{adjustbox}
\label{table:encounterRecord}
\vspace*{-0.3cm}
\end{table*}

\begin{table*}[h!]
\centering
\caption{NetFlow example}
\begin{adjustbox}{max width=\textwidth}
  \begin{tabular}{*{10}{|c}|}
\hline
  Start time & Finish time & Duration & Source IP & Destination IP & Protocol& Source port & Destination port & Packet count & Flow size  \\
\hline
1334252579.845 & 1334252599.576 & 19.731 & 173.194.37.7 & 10.15.225.126 & TCP & 80 & 60385 & 223 & 224862\\
\hline
\end{tabular}
\end{adjustbox}
\label{table:netflow}
\vspace*{-0.3cm}
\end{table*}

\section{Mobility encounters}
Pairwise mobile encounter events provide opportunities for dissemination events such as content dissemination \cite{enMetric, elsherief2017novel} and infection spreading through direct encounter \cite{inTrace}.

Consequently, designing effective content distribution, routing schemes and infection tracing back approaches require encounter understanding and realistic modeling. 
While encounter events have been analyzed in several previous studies (e.g. \cite{enNodal,enMetric}), here we develop new insights into pairwise events by considering the following: 
1) \emph{Device types}: We distinguish between encounters among the three groups in our analyses (\emph{FF}, \emph{CC}, \emph{FC}).
2) \emph{Large-scale data}: The data is first of its kind in terms of its size where it covered more than 140 buildings with different categories. Also, we analyze mainly indoor (in-building) encounters, unlike most previous studies. 
3) \emph{Traffic-encounter analysis}: Daily encounter patterns at buildings are analyzed per device type, then their correlation to traffic patterns are studied for the first time
in Section \ref{sec:pairwise}. 

\subsection{Daily Encounter Duration at Buildings}
\label{sub:encDurFvsC}
The pairwise statistical summary of mobility encounters are generated from daily encounter records at each building.\\
The total encounter duration, \textit{E}, of a pair of users $(u1,u2)$ during day \textit{d} at building \textit{B}, $E_{Bd}(u1,u2)$ is computed as:
$E_{Bd}(u1,u2) = Sum \sum_{i=1}^{i=n} E_{Bd}(u1,u2)_{i}$,
where $n$ is the number of encounters, and $E_{Bd}(u1,u2)_{i}$ is the duration of encounter $i$ between $u1$ and $u2$ on day $d$ at building $B$, respectively. If a pair of users encounter again on a different day or in a different building, that encounter is considered separately. Overall, $\approx20\%$ have encountered at least twice in any building. \\
The pairs are then separated based on their pair device types.  
\\
\begin{table*}[h!]
\centering
\caption{Daily Encounter Duration in Seconds}
\begin{adjustbox}{max width=\textwidth}
  \begin{tabular}{*{8}{|c}|}
\hline
\textbf{Pairs-Types}	&	\textbf{Min.}	&	\textbf{1st Qu.}&	\textbf{Median}	&	\textbf{Mean}	&	\textbf{3rd.Qu.}	&	\textbf{Max.}	&	\textbf{Std.}	\\
\hline
Flute-Flute	(\emph{FF}) &	1	&	27	&	84	&	528.2	&	367	&	77170	&	1417	\\
\hline
Cello-Cello (\emph{CC})	&	1	&	135	&	844	&	2061	&	2834	&	84580	&	3244	\\
\hline
Flute-Cello	(\emph{FC}) &	1	&	34	&	169	&	954.4	&	855	&	80660	&	2021	 \\
\hline
\end{tabular}
\end{adjustbox}
\label{table:EnDurationSummary}
\vspace*{-0.3cm}
\end{table*}

\begin{figure}[ht!]
	\centering
	\includegraphics[width=0.45\textwidth]{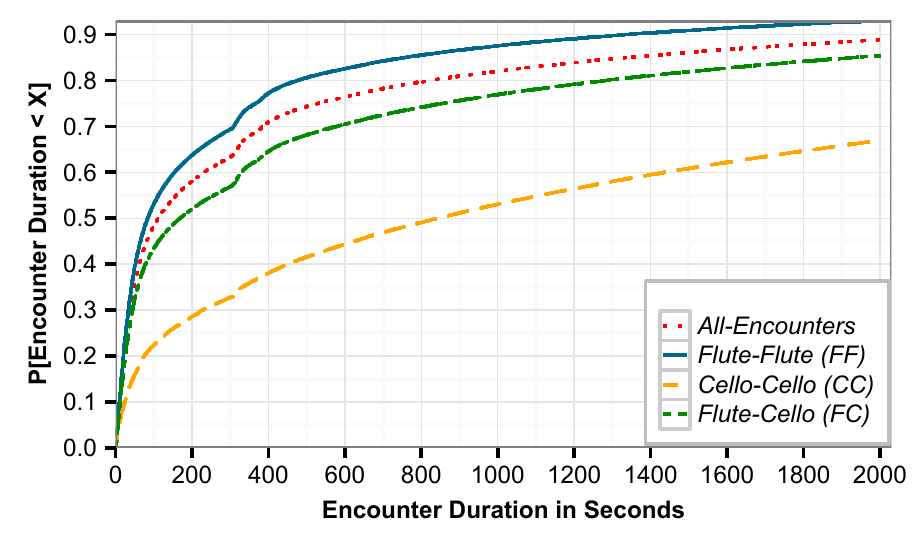}
	\caption{Encounter duration CDF based on encounter pair device type.}
	\label{cdfDur}
\vspace*{0.2cm}
\end{figure}

The daily encounter duration based on device types are summarized in Table \ref{table:EnDurationSummary}. 
From the table, it is clear that \emph{CC} pairs have longer encounter duration than other kinds of pairs. For example, the mean \emph{CC} daily encounter duration is \textbf{290\% longer} than the \emph{FF} pairs. This result is beneficial when modeling the encounters based on device type, or with applications that use the encounter duration.\\   
  Figure \ref{cdfDur} shows the CDF of the encounter duration for 95\% of pairs (the highest 5\% is omitted for clarity). Note that 80\% of \emph{FF} encounters have daily encounter duration \textbf{$\leq$8 minutes}, while only 40\% of \emph{CC} encounters are $\leq$8 minutes.
For all encounters, 33\% are $\leq$38s, dubbed \textit{short}, the next 33\% are $\leq$317s, called \textit{medium}, and $>$317s are \textit{long} encounters.
This definition will be used in Sec. \ref{sec:pairwise} for pairwise analysis of the correlation between encounter duration and traffic profile similarity.

\subsection{Encounter Duration Statistical Distributions}
Eleven distributions are fit to the total daily encounter duration using maximum likelihood,  and goodness-of-fit test methods: Power-law (\emph{Pl}), Weibull (\emph{W}), Gamma (\emph{G}), Lognormal (\emph{Ln}), Pareto (\emph{Pr}), Normal (\emph{N}), Exponential (\emph{Ex}), Uniform (\emph{U}), Cauchy (\emph{C}), Beta (\emph{B}) and Log-logistic (\emph{Ll}). 
The Kolmogorov-Smirnov (\emph{KS}) statistic is used to evaluate the distributions fitness. The three best-fit distributions are selected and presented in table \ref{tab:avgEdur}. For example, 74\% of the buildings have \textbf{power-law} (\emph{Pl}) distribution as the best fit for their \emph{FF} pair daily encounter duration, while only 39\% of buildings have \emph{CC} daily encounter distribution following power-law.
 Also, \textbf{table \ref{tab:avgEdur}} shows the percentage of buildings that have KS-statistic with less than a threshold, specifically $\leq 5\%$ and $\leq 10\%$. This is calculated to see if there is a distribution that can be a good fit for the majority of buildings, even if it is not the first-best fit. 
Power-law and log-logistic distributions usually have KS-test with $\leq$10\% for 92\% for \emph{FF} and \emph{FC} pairs.  

\begin{table*}[tb]
  \centering
  \caption{Best fit distributions for total daily encounter pairs duration based on pairs classifications. Percentage of buildings shown in brackets.
 (\textbf{Pl}: Power Law, \textbf{Ll}: Log Logistic, \textbf{W}: Weibull,  \textbf{G}: Gamma,  \textbf{Pr:} Pareto,  \textbf{Ln}: Log Normal, \textbf{B}: Beta).} 
 
  \label{tab:avgEdur}
  \begin{tabular}{||c|c|c|c||c|c||}
  \hline
 \textbf{Pair-Types}&\textbf {1st best fit} & \textbf{2nd best fit} & \textbf{3rd best fit}&\textbf{KS-test<=5\%}&\textbf{KS-test<=10\%}\\
 \hline
 \hline
 Flute-Flute & Pl[74\%], Ll[10\%],	&	    
 
 Ll[29\%], Ln[26\%],	&

Ll[39\%], Pr[24\%],&
         Pl[72\%], Ll[19\%]  &
         Pl[92\%], Ll[92\%]
         
               \\
           (\emph{FF})  &     Pr[8\%]           &
                     Pr[19\%], W[11\%]   &
                       Ln[21\%]    &
             Ln[15\%], Pr[15\%]     &
             Pr[89\%], Ln[85\%]
                       \\
  
 \hline 
    Cello-Cello & Pl[39\%], W[21\%],	&	 B[22\%], Ll[18\%]	&
    Ll[25\%], W[24\%],&
    Pl[34\%], W[32\%]         &
    Ll[87\%], Pl[86\%]
    \\
     (\emph{CC}) &  B[14\%]	& Pl[16\%], W[14\%]	&
        Pr[13\%], Pl[12\%]&
    B[30\%], G[16\%]&
    W[78\%], Ln[71\%]
    \\
   
   \hline   
   
    Flute-Cello & Pl[67\%], Ll[13\%],	&	Ln[28\%],   Ll[20\%], &	Ll[46\%], Ln[16\%]&
   Pl[61\%], Pr[12\%]&
   Pl[92\%], Ll[92\%]\\
       (\emph{FC})  &  Pr[9\%]	&	
    Pr[19\%], W[18\%], Pl[10\%]	&	
    Pr[14\%]& 
    Ll[11\%], W[9\%]&
    Pr[80\%],Ln[78\%]
    
    \\
    
   \hline   
   
\end{tabular}
  
\end{table*}

\section{Web traffic profile}
We use \textit{NetFlow} traces to analyze traffic behavior of user devices.
In \cite{Alipour2018}, we analyzed traffic on an \textit{individual} level.
We found \textit{cellos} to generate 2x more flows than \textit{flutes}, while the \emph{flute} flows are 2.5x larger.
Also, flow sizes were found to follow a Lognormal distribution, while flow inter-arrival times (\emph{IAT}) follow beta distribution, with high skewness/kurtosis,
hinting at infrequent extreme values (e.g., \emph{flutes} incur more extreme periods of inactivity, caused by higher mobility).
In this study, we conduct \textit{pairwise} (vs. \emph{individual}) level analysis of mobility and traffic.

To analyze traffic patterns of users for all buildings and days, 
we first define a \textbf{\emph{Traffic Profile (TP)}} for each user based on \emph{NetFlow} traces. This traffic profile is efficient to calculate and granular enough for our analysis:

\begin{itemize}
\item First, we select a set of popular websites for analysis based on total bytes sent and received, filtering out websites with little usage. 
The IP address of selected websites form the dimensions of the traffic vector, denoted as $\kappa$.
There are $\approx10,000$ IP addresses in $\kappa$, with average \textit{daily} traffic from few $MB$s, and up to $690GB$s.

\item Next, for each address, $IP_j$, we calculate $B_{(i,j)}$,
defined as the natural logarithm of total traffic \textit{user i}, $U_i$, has sent to or received from $IP_j$.
This forms the initial \emph{traffic vector} for $U_i$ , consisting of  $B_{(i,j)} , \forall j \in \kappa$.

\item Finally, we apply \emph{term frequency-inverse document frequency} (\textbf{\emph{TF-IDF}} \cite{salton1986introduction}) to the collection of traffic vectors of all users. This reduces the effect of wildly popular websites, and identifies websites that can distinguish between users' online behavior, enabling us to study the richness in the access patterns.
In this context, each $IP_j$ is a \emph{term} and each user \emph{traffic vector} is a \emph{document}.
\emph{TF-IDF} is calculated as the product of term frequency 
(the number of times a term appears in a document, corresponding to $B_{(i,j)}$ in our context, which reflects the bytes $IP_j$ exchanged with $U_i$),
and inverse document frequency (the inverse of number of documents (\textit{users}) the term (\textit{IP}) occurs in) \cite{idf}.
Each row of the resulting matrix is a \textit{traffic profile, $TP_i$,} of user $U_i$, as depicted in Fig \ref{fig:tfidf}.

\begin{figure}[ht!]
    \centering
    \[
    \begin{array}{cccc}
    & \text{$\text{IP}_{j}$}\\
    \text{\emph{TF-IDF}=} &
      \begin{bmatrix}
          & \vdots  &  \\
         <\cdots &B_{i,j} &\cdots>\\
         & \vdots  & 
      \end{bmatrix} &
      \text{$\leftarrow$ $TP_i$}
      \end{array}
    \]
    \caption{TF-IDF Matrix: Each row is a user profile.}
    \label{fig:tfidf}
\end{figure}

\end{itemize}

This process is applied for \emph{NetFlow} data of \textit{every} building on \textit{each day}, to enable spatial (across buildings) and temporal (across days) analysis of user traffic profiles\footnote{If a building on a specific day has less than 20 encountered pairs, that (building, day) pair is omitted, to maintain statistical significance.}.
For pairwise comparison of traffic profiles, we use Cosine similarity which computes the cosine of the angle between two user profiles.

\section{Pairwise encounter-traffic relationship}
\label{sec:pairwise}
With mobility encounters and traffic profiles as the pillars, here, 
we take steps to investigate \textit{"whether physical encounters are correlated with the similarity of traffic profiles"}.
This analysis outlines our initial findings in the \textit{pairwise} (\textit{encounter}) dimension of mobility-traffic analysis, 
following our work in \cite{Alipour2018} that focused on \textit{individual} aspect of combined mobility-traffic modeling, and providing the foundation for \textit{collective} (\textit{group}) analysis in the future. We start with simple steps, and increase the complexity of methods gradually.

As a first step, we seek to establish whether the traffic profiles of encountered pairs are more similar compared to traffic profiles of non-encountered pairs of users.
For this purpose, we calculate $enc = Sum \sum_{\forall (i,j) \in Z}{\frac{cosine(TP_{i}, TP_{j})}{|Z|}}$. Here, $Z$ denotes the set of all encountered pairs of users.
Similarly, for all non-encountered pairs, $Z'$, $nonenc = Sum \sum_{\forall (i,j) \in Z'}{\frac{cosine(TP_{i}, TP_{j})}{|Z'|}}$.
This calculation is carried out on each building every day.
Overall, we observe that \textbf{$\approx$93\% of the time \textit{enc > nonenc}}, with
the main exceptions being buildings close to bus stop hubs on campus, with a high pass-by rate of users; resulting in many short encounters that do not show traffic similarity.
With that simple observation, next, we asked whether the difference between traffic similarity of encountered and non-encountered is statistically significant.
\textit{Mann–Whitney U} test \cite{mann1947test} was applied on the two independent sets, with the null hypothesis being the two sets are drawn from the same distribution.
We find that for $86\%$ of (building, day) tuples, we can reject the null hypothesis ($p < 0.05$).
This shows that \textbf{\textit{in most cases, the traffic profiles of encountered pairs are more similar and the difference between the two groups is statistically significant}}.
Logistic regression analysis shows that similarity of traffic profiles is significantly associated with the probability of encountering for all days in several buildings, such as the computer department with a big user base, for $\approx90\%$ of days in libraries, and only for $\approx48\%$ of days in gym and recreation centers where users normally do not use networks as much.
The next question is how consistent these differences are across: 
\begin{enumerate}
\item Device type categories: 
As discussed earlier, usage patterns of devices differ based on form factor (e.g., on-the-go \textit{flutes} vs. sit-to-use \textit{cellos}). 
We compare flute-flute (\emph{FF}),  cello-cello (\emph{CC}), and flute-cello (\emph{FC}) encounters.
\item Weekday vs. Weekend: 
We established significant differences between mobility and traffic patterns of weekdays and weekends in \cite{Alipour2018}. Here we analyze the mobility encounter-traffic interplay across the weekdays and weekends.
\item Encounter duration: 
We define three encounter duration categories using 3 bins of equal frequency: short ($<0.6 min$), medium ($0.6-5 min$) and long ($>5 min$). 
We then analyze each group for correlation between encounter duration and traffic profile similarity.
\end{enumerate}

\subsection{Device type categories}
\label{sub:devTypes}
\begin{figure}[ht!]
	\centering
	\includegraphics[width=0.49\textwidth]{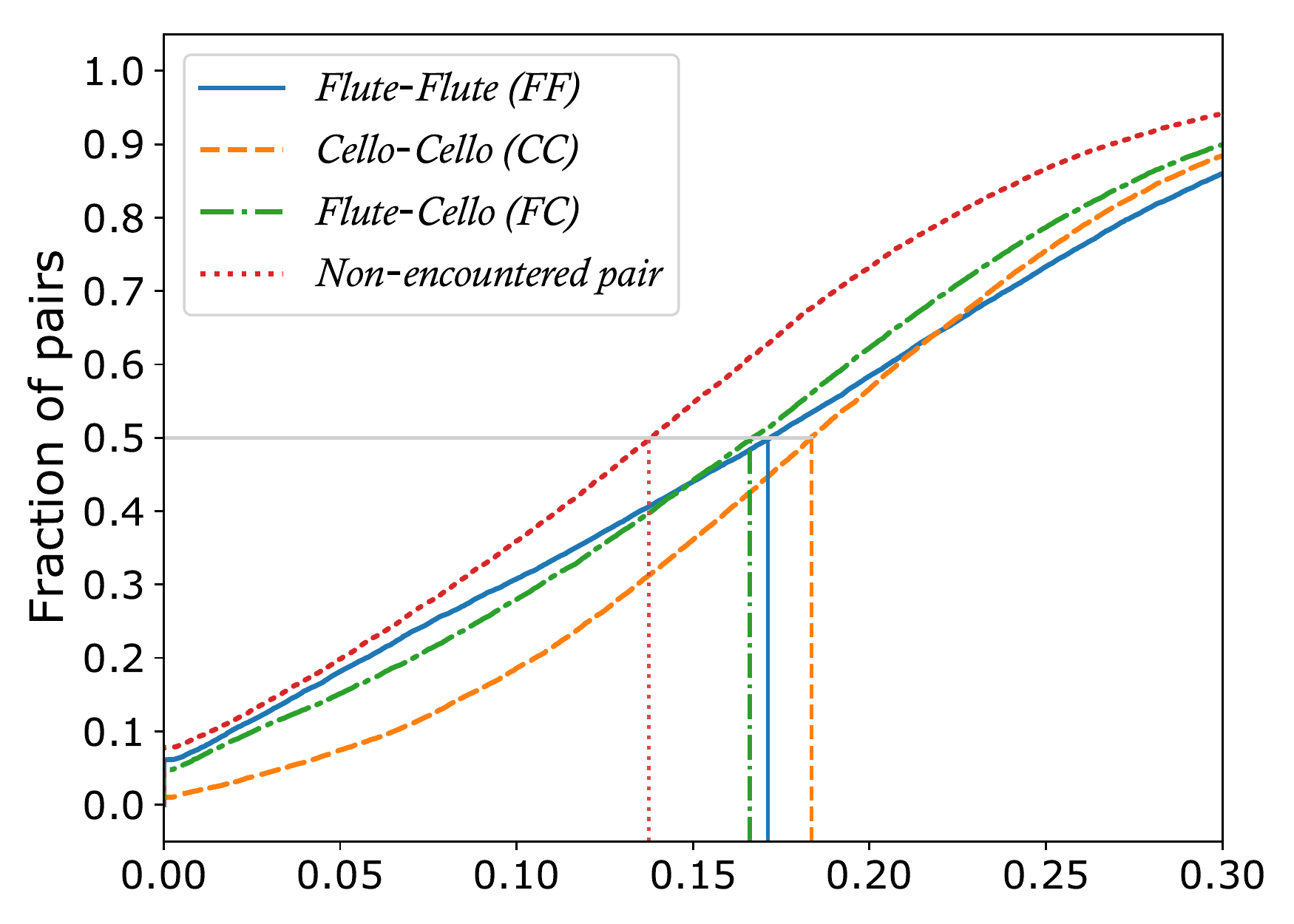}
	\caption{CDF of pairwise cosine similarity of traffic profiles across device types (vertical lines denote medians).}
	\label{fig:cdf_devType}
\end{figure}

We analyze how similarity of traffic profiles for encountered pairs varies when two \textit{flutes} meet (\textit{FF}), two \textit{cellos} meet (\textit{CC}) or a \textit{flute} meets a \textit{cello} (\textit{FC}).
The results, as presented in Figure \ref{fig:cdf_devType}, show that the similarity of \textit{CC} is slightly higher than the other groups, while the \textit{FF} and \textit{FC} groups show similar trends.
Notably, however, \textbf{\textit{all three encountered groups are significantly different from the non-encountered group}} ($p < 0.05$).
This is consistent across most buildings.
Given the context of the traces, we suspect heavy use of laptops for educational content on campus.
Further analysis website content may shed light on the shared interests among encountered users with various forms of devices.
We leave this for future work.

\subsection{Weekday vs. Weekend}
\begin{figure}[ht!]
	\centering
	\includegraphics[width=0.49\textwidth]{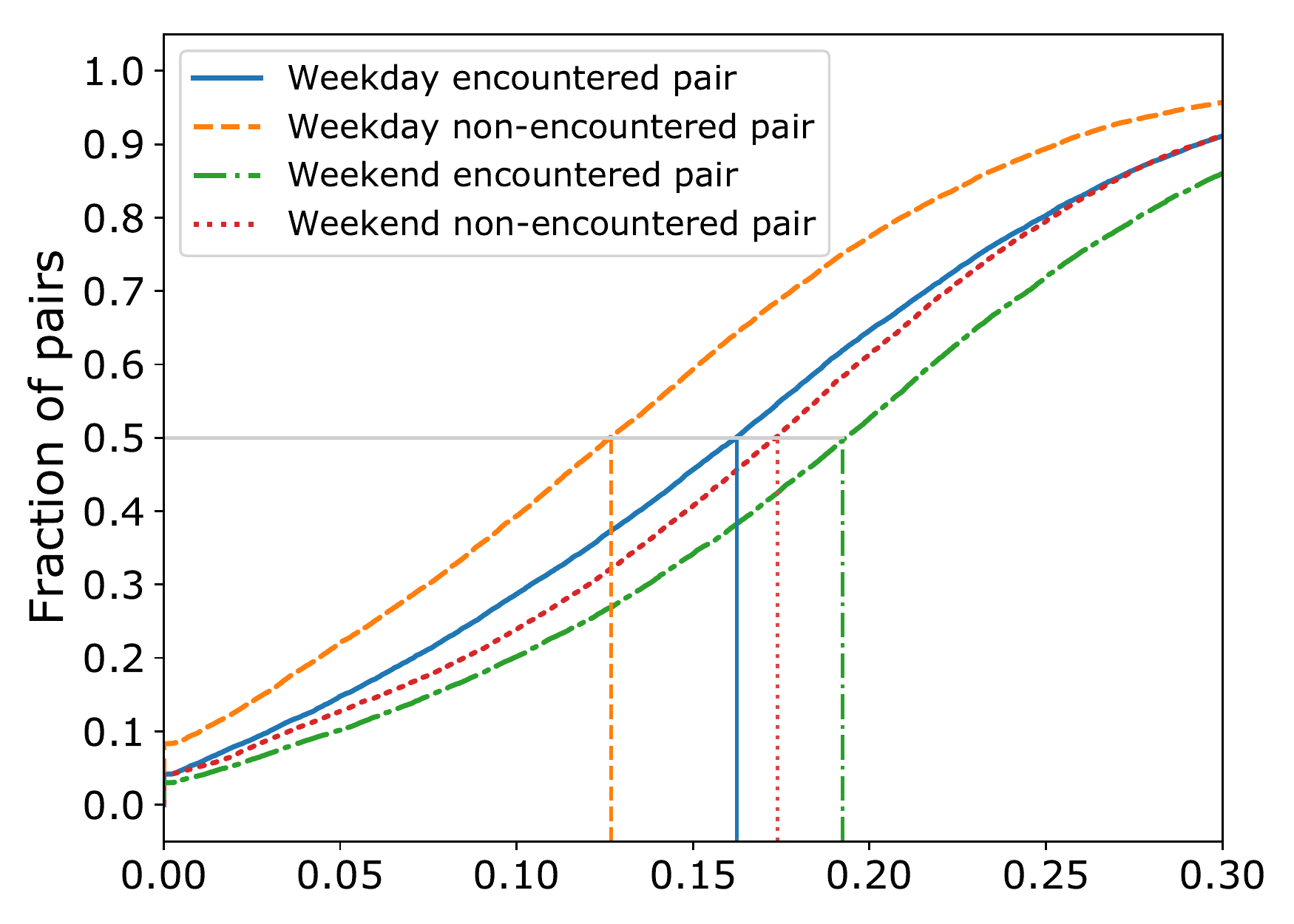}
	\caption{CDF of pairwise cosine similarity of traffic profiles in weekdays and weekends (vertical lines denote medians).}
	\label{fig:cdf_wdwe}
\end{figure}

Intuitively, there are significant differences between weekdays and weekends in user behavior, and consequently their mobility and traffic patterns. In \cite{Alipour2018}, we found that numbers of user devices on campus drops significantly on weekends, but the remaining devices do not show significant differences in terms of flow size, packet count, and active duration. Here we \emph{identify and quantify the encounter-traffic correlation over weekdays/weekends} for the first time.
Results are depicted in Fig \ref{fig:cdf_wdwe}. We find that the \textbf{\textit{pairwise similarities of weekend pairs to be overall higher than their weekday counterparts}} regardless of an encounter (or not), with weekend \textit{non-encountered} pairs being more similar than weekday \textit{encountered} pairs.
This is explained by observing significantly reduced mobility of devices on weekends. 
For example, median radius of gyration for \emph{cellos} drops by 66\%, and by 15\% for \emph{flutes} \cite{Alipour2018}. 
In addition to decreased mobility, most activity is clustered around several academic buildings with research labs (33\% of APs handle no \emph{flute} traffic on weekends, while 56\% receive no \emph{cello} traffic).
Thus the increase in traffic similarity
during weekends might be explained by the presence of researchers collaborating on related fields of interest and accessing similar content.

\subsection{Encounter duration}
\begin{figure}[t!]
	\centering
	\includegraphics[width=0.49\textwidth]{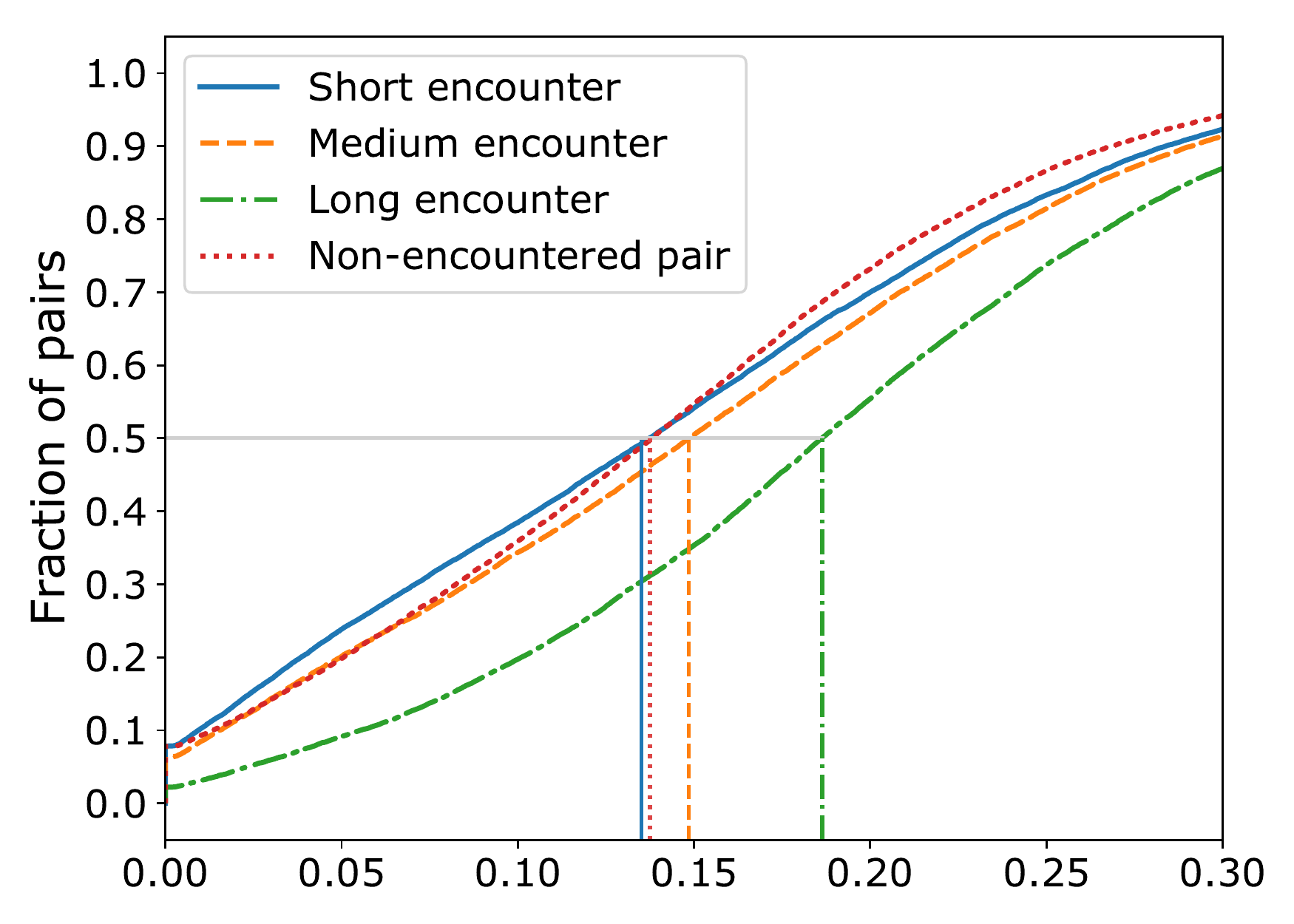}
	\caption{CDF of pairwise cosine similarity of traffic profiles with different encounter durations (vertical lines denote medians).}
	\label{fig:cdf_encDurCat}
\end{figure}

Based on the encounter durations introduced in Section \ref{sub:encDurFvsC}, we define three encounter duration categories with 3 bins of equal frequency: short ($<0.6 min$), medium ($0.6-5 min$) and long ($>5 min$). 
As depicted in Fig. \ref{fig:cdf_encDurCat}, the short encounter group is not significantly different from the group of non-encountered pairs ($p > 0.05$).
However, the differences between the other groups are statistically significant, with the \textbf{\textit{long encounter group showing the highest similarity of traffic profiles}}, hinting at a correlation between the duration of encounter and traffic profile similarity.

\begin{figure}[ht!]
	\centering
	\includegraphics[width=0.49\textwidth]{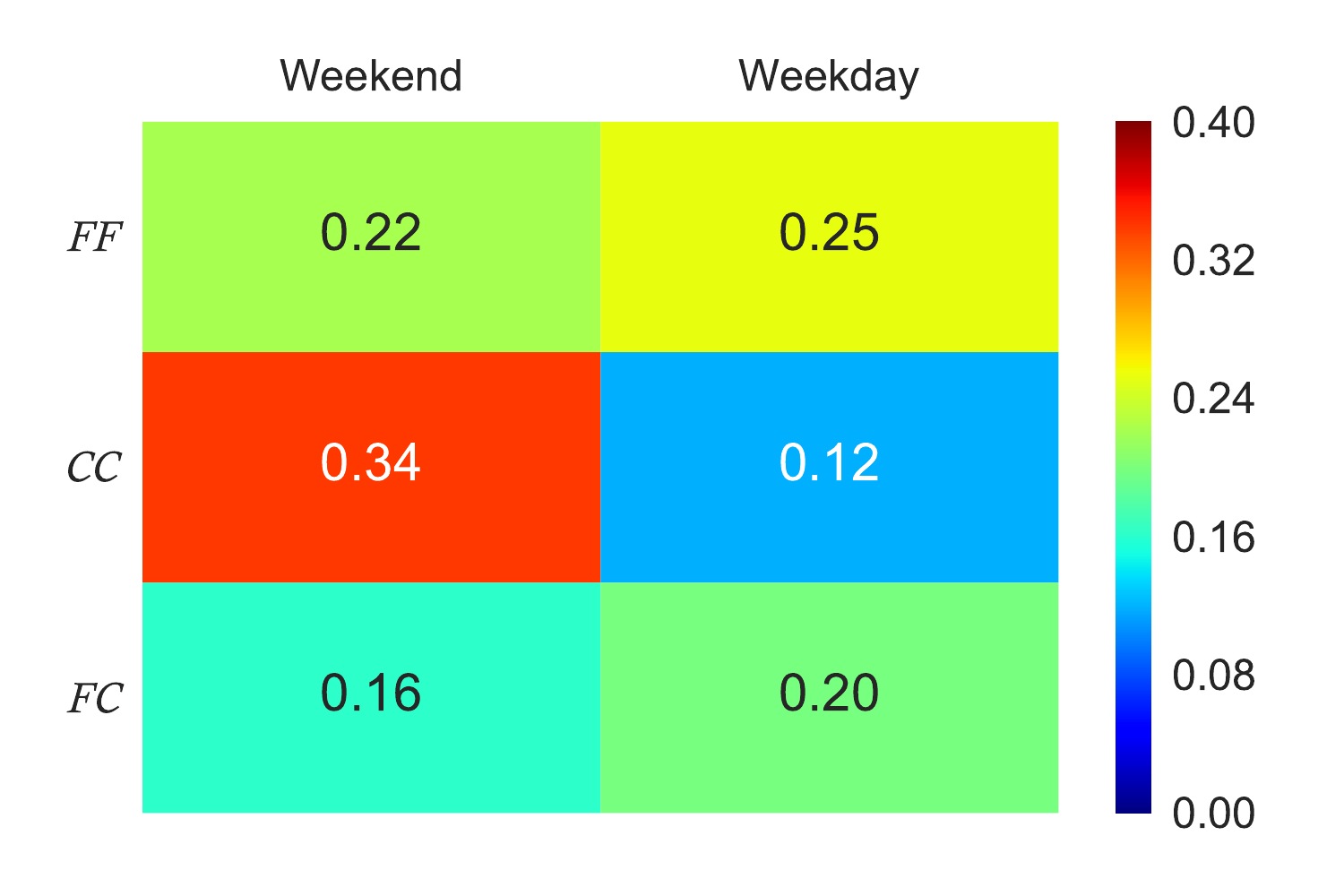}
	\caption{Pearson correlation coefficient between encounter duration and traffic profile similarity. (\emph{Cello-Cello} (\emph{CC}) encounters on weekends show the highest positive correlation.)}
	\label{fig:heatmap_corr_devType_wdwe}
\end{figure}

Hence, we investigate this correlation.
We found insignificant correlation for short and medium encounters (based on \emph{Pearson} and \emph{Spearman} correlation coefficients), 
however there is a \textit{small positive} linear correlation between encounter duration and traffic profile similarity for long encounters ($\rho=0.21$).
Breaking down the correlations into different device types and weekday/weekend (Fig. \ref{fig:heatmap_corr_devType_wdwe}), shows the highest correlation for \emph{Cello-Cello} (\emph{CC}) encounters on weekends, supporting our earlier observation.

\begin{figure}[ht!]
	\centering
	\includegraphics[width=0.5\textwidth]{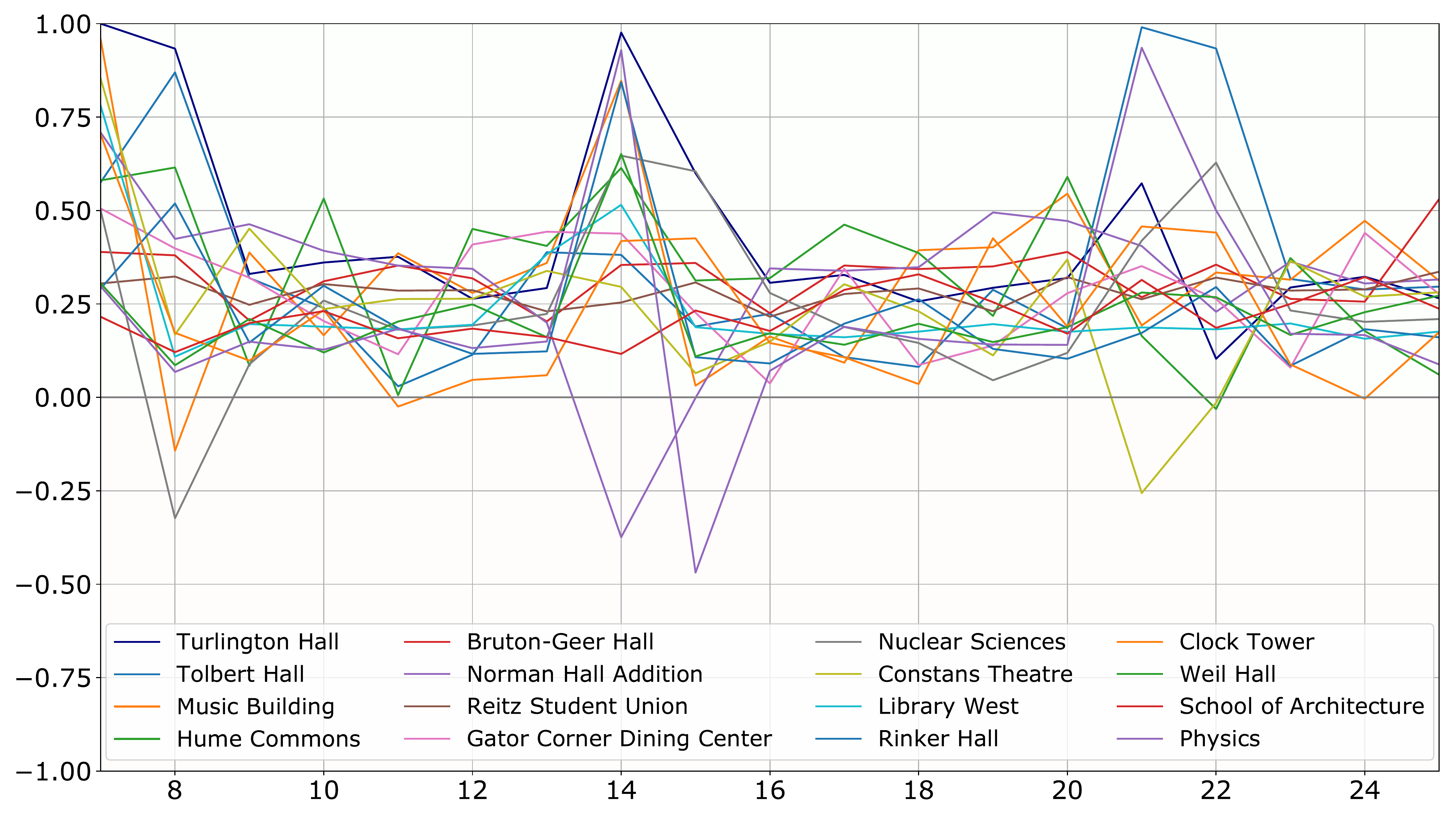}
	\caption{Pearson correlation coefficient between encounter duration and traffic profile similarity for different buildings across days (X-axis).}
	\label{fig:daily_corr}
\end{figure}

Overall, the correlation between encounter duration and traffic profile similarity is dynamic, changing across space and time. 
Fig. \ref{fig:daily_corr} shows a time-series plot of the linear correlation coefficient of several buildings on campus for more than 3 weeks.
It shows how the correlation varies across time in different buildings, with rapid changes every 7 days (around weekends). 
Surprisingly, a few buildings show significant \textit{negative} correlations on weekends (e.g., music and theater buildings), while others show significant \textit{positive} correlations on the same days (mostly academic buildings).
Further analysis of the other buildings and its interaction with mobility encounters and traffic profiles are left for future work.

\section{Learning encounters}
\label{sec:dnn}
Given the relationships shown so far, there seems to be great potential in training a \textit{machine learning model} that can learn to predict an encounter given two traffic profiles.
Such a model has several practical applications.
Given two traffic profiles, if it is possible to \textbf{predict} they have encountered on a certain day in a building with good accuracy, then there is useful information in the relationship of mobility encounter and traffic profile similarity, which could be used in design of encounter-based \textit{dissemination} protocols, analysis of \textit{rumor anatomy}, or \textit{tracing} of disease spread \cite{inTrace} even if mobility traces are not reliable for each user (for example due to MAC address \textit{randomization}), but traffic profiles of users are accessible (via \textit{authentication} mechanisms identifying users at a higher OSI layer).

To investigate the feasibility of this task, for every pair of users, their \textit{traffic profiles} in each building and on each day are coupled as input (either through concatenation or taking the absolute differences, with the former depicted in results and figures), and a binary target label is assigned based on whether the pair has encountered on that day and building.
Since most pairs of users do not typically encounter on a day, predicting a negative label is rather trivial in this case. To prevent this bias, we sample this dataset to make sure each label is represented by an equal number of samples for our models.

\subsection{Random Forest}
We first used Random Forests \cite{ho1995random,Breiman2001} for this classification task, which is a well-established algorithm used for supervised learning problems.
Our work showed that on a building (the computer department) the algorithm achieved a promising \textbf{$\approx70\%$} accuracy on average across all days, based on stratified k-fold cross-validation, without employing any preprocessing or parameter tuning of the model.
Next, we applied a dimensionality reduction algorithm, using \textit{Singular-Value Decomposition (SVD)} to preprocess the input vector. 
This technique is adapted from Latent semantic analysis of natural language processing.
Its application improved the accuracy, in the same settings, to \textbf{$\approx73\%$}.
This lead to the idea of using stacked auto-encoders (SAE) to retain information and connect the SAE to a \textbf{\textit{deep, fully-connected neural network (DNN)}} for classification.

\begin{figure}[ht!]
	\centering
	\includegraphics[width=0.5\textwidth]{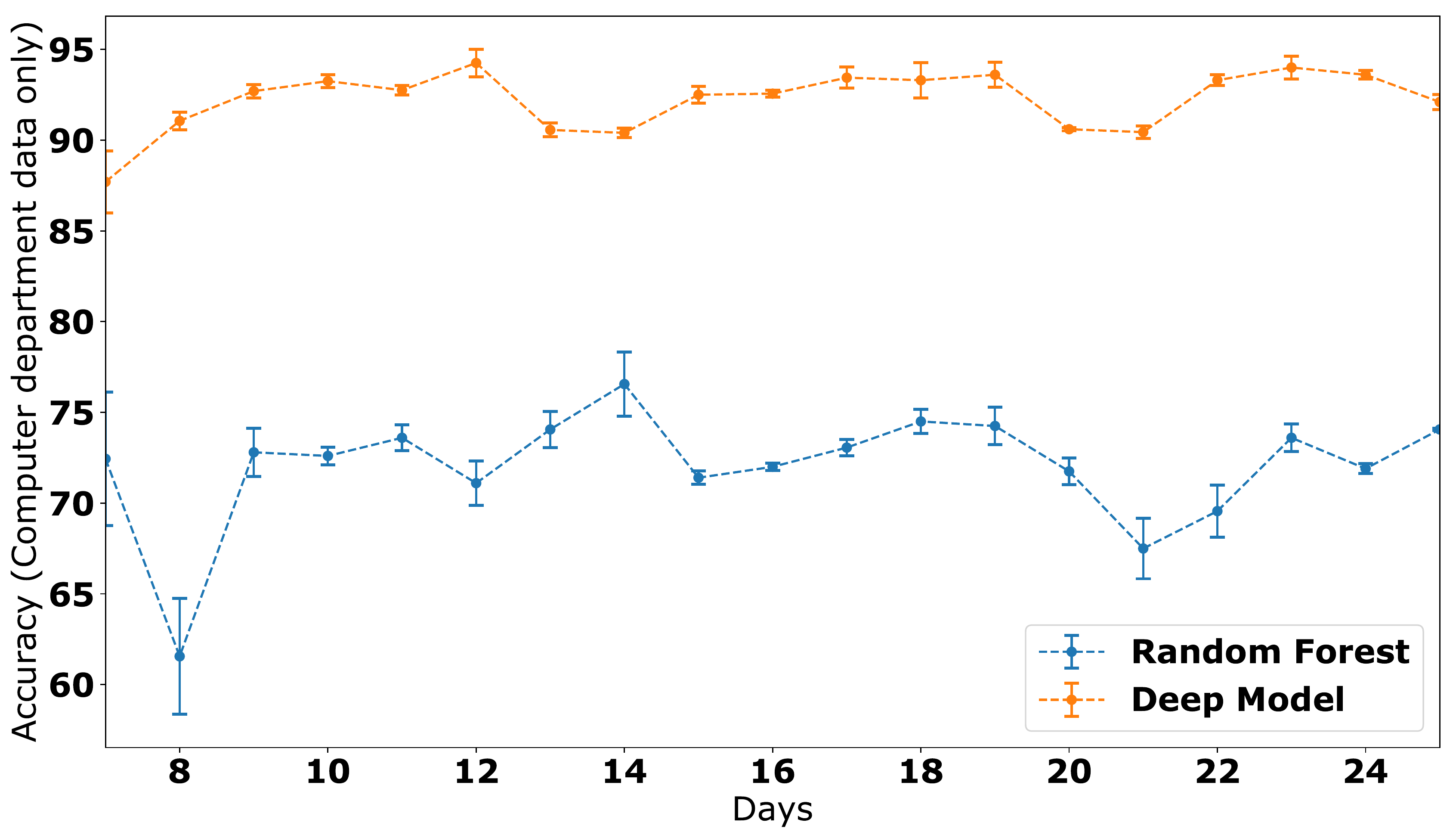}
	\caption{Accuracy of random forest and deep learning model for encounters in the computer department across days (X-axis).}
	\label{fig:accuracy_days_CISE}
\end{figure}

\subsection{Deep learning}
We utilized several recent ideas from the field of artificial intelligence to improve our learning of encounters significantly.
Auto-encoders are a class of artificial neural networks that are trained in an unsupervised fashion to learn an efficient representation of their input.
In simple terms, the network consists of two stages: encoder and decoder. The encoder consists of layers of decreasing size, that is then connected to the decoder, which is made up of layers of increasing size. The objective is to reconstruct the input as accurately as possible with purely unsupervised learning.
Stacked auto-encoders (SAE) have been used in various applications to extract features \cite{masci2011stacked}, reduce dimensionality \cite{nowicki2017low}, as well as denoising the corrupted variations of input \cite{vincent2010stacked}.

\begin{figure}[ht!]
	\centering
	\includegraphics[width=0.5\textwidth]{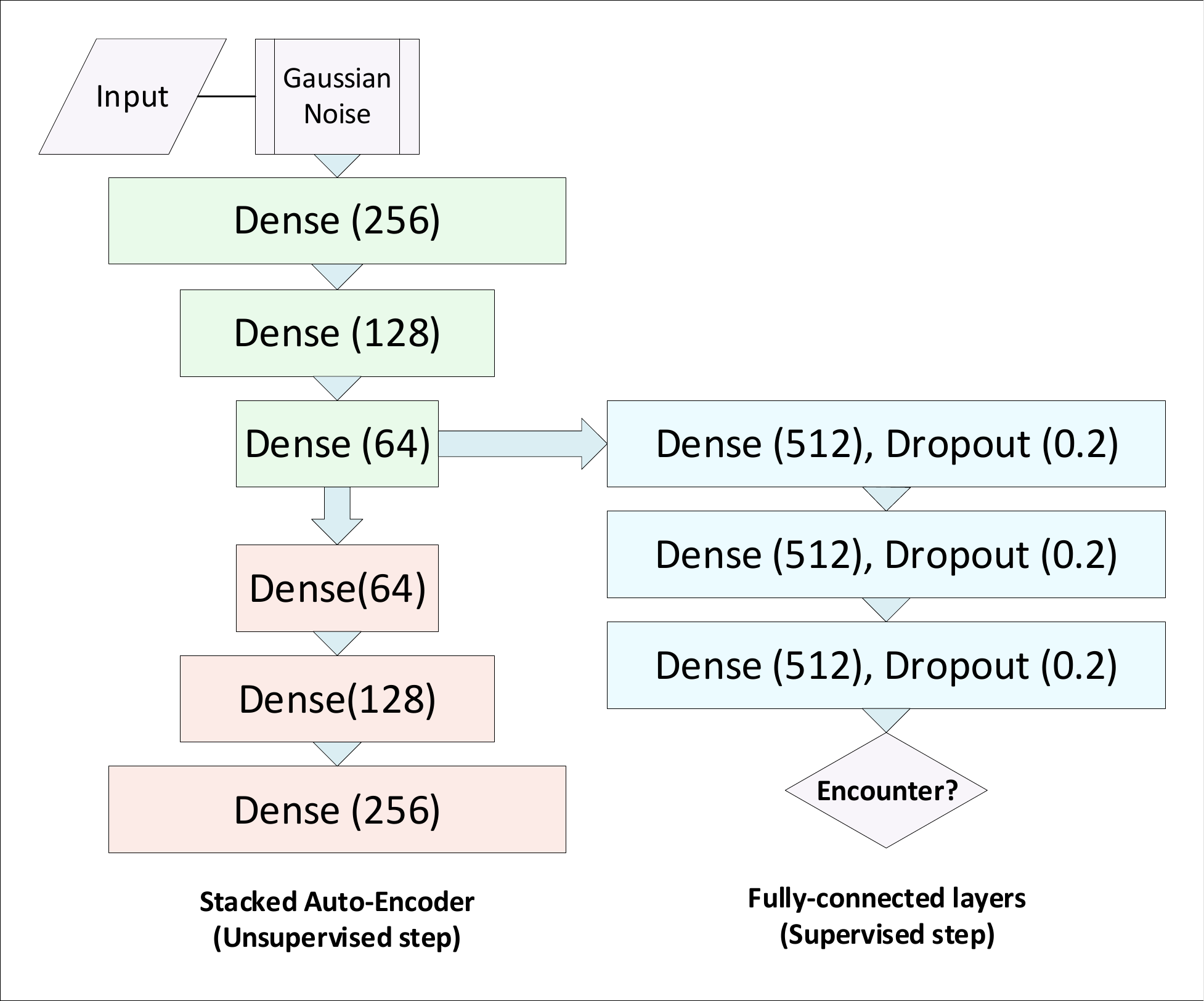}
	\caption{Architecture of the deep learning model. Numbers show the number of neurons in each layer (internal details omitted for brevity).}
	\label{fig:dnn}
\end{figure}

We use a \textit{Stacked De-noising Auto-Encoder (\textbf{SDAE})} \cite{vincent2010stacked}. This network is provided with input data corrupted by Gaussian noise, and is trained to reconstruct the original, uncorrupted input similar to a traditional auto-encoder. Thus, denoising becomes a training criterion, while dimensionality reduction of input is also achieved (We kindly refer the reader to \cite{vincent2010stacked} for details of SDAE and comparison with SAE).
Then, the encoded data points are fed to a fully connected, multi-layer neural network.
Comparing the results to the random forest, there is a significant \textbf{increase} in accuracy to an average of \textbf{\textit{92\%}} for the same building and days as the random forest classifier.
Comparing \textit{device type categories} of encounters, we find \textit{cello-cello} encounters to be the most distinguishable, followed by \textit{flute-cello} and \textit{flute-flute}. However, the difference between accuracies for different device type categories is <5\% in most locations and dates, a testimony to the \textit{robustness} of the model.
This accuracy is also stable across time, with the median of accuracy, for \textit{weekdays} at 93.25\%, and \textit{weekends} at 90.75\%, for the computer department samples.
The much higher accuracy comes at the cost of high compute power costs and complexity of the model.
\textbf{Fig. \ref{fig:accuracy_days_CISE}} shows the results of both the \textbf{random forest} and the \textbf{deep neural network model} for this building across $\approx3$ weeks.
We used \textit{stratified k-fold cross-validation}, \textit{early stopping} and \textit{dropout} layers to regularize the network and alleviate overfitting. An illustration of the architecture is presented in Fig. \ref{fig:dnn}.

\section{Conclusions and future work}
\label{sec:conclusion}
In this study, we present the first steps to analyze and quantify the relation between mobility and traffic.
Focusing on the pairwise (encounter) dimension of mobility, its interplay with the traffic patterns of mobile users was studied.
This work has implications for realistic modeling and simulation, offloading through opportunistic encounters, as well as implementation and benchmarking of encounter-based services such as content sharing, mobile social networks and encounter-based trust.
We use extensive, highly granular datasets (30TB in size), in more than 140 buildings on a university campus, including information about WiFi associations, DHCP and NetFlow, covering the dimensions of mobility and network traffic. 

To answer our main question of ‘How do device encounters affect network traffic patterns, across time, space, device type, and encounter duration?’,
We analyze mobility encounters and presented their statistical characteristics.
We defined traffic profiles and utilized numerical statistics and machine learning techniques.
Power law and Log logistic distributions, fitted to daily encounter duration, have KS-test $\leq$10\% in 92\% of buildings for Flute-Flute (FF) and Flute-Cello (FC) encounters, and 86\% for Cello-Cello (CC) encounters. Also, \textit{CC} pairs have longer daily encounter duration. 
Analyzing traffic, we find significant differences between traffic profile similarity of encountered versus non-encountered pairs for device type categories (\textit{FF}, \textit{CC}, \textit{FC}), with the highest similarity being the CC group.
Further, comparing weekdays and weekends, in both cases, the encountered pairs are more similar, with the distinction that weekend traffic profiles are more similar than weekdays'.
Analysis of correlation between encounter duration and traffic profile similarity revealed short and medium encounters not being significantly different from the non-encountered group, while the long encounters show significantly higher similarity.
We also employed random forests and created a deep neural network (DNN) model to predict encounters of pairs of user traffic profiles, with a very high accuracy (up to 94\% depending on settings).

The findings in this paper are not currently captured by any of the existing mobility or traffic models, while having important implications in many contexts, such as predictive caching, information dissemination, opportunistic social networks and infection tracing. This provides a compelling case for integrated traffic-mobility models that consider multiple dimensions of social context (individual, pairwise, and group).
We plan to further investigate the causal relationship between mobility and traffic for pairwise and \textit{collective} (group) dimensions in the future. Exploration of applications of our learning methods, its privacy implications, and potential improvements are also left for future work.

\section*{Acknowledgement}
This work was partially funded by NSF 1320694, and Najran University, Saudi Arabia.
We gratefully acknowledge the support of NVIDIA Corporation with the donation of the Titan Xp GPU used for this research.

\nocite{*} 

\bibliographystyle{IEEEtran}
\bibliography{main}

\end{document}